# *Socioduality:* A Relational Process Framework for Human–AI Interaction

Mehmed Zahid Çögenli, PhD
Department of Occupational Health and Safety, Faculty of Health Sciences
Uşak University, Uşak, Türkiye
mzahid.cogenli@usak.edu.tr | ORCID: 0000-0003-3018-4157

## Abstract

Human-AI research often evaluates individual capabilities, combined performance, or final outputs. These approaches do not preserve how one party's response becomes part of the conditions under which the other party's next contribution is formed. This article introduces socioduality, defined as a sequential, reciprocal, and history-carrying relational process between two distinguishable parties in which a response from one party becomes part of the observable conditions under which the other party's subsequent contribution, judgement, decision, or action is formed. The construct is specified for human-AI dyads and operationalised through nested units: moves, confirmed sociodual episodes, linked pathways, and the broader interaction container. A minimum episode $A_1 \rightarrow B_1 \rightarrow A_2$ requires evidence of both response contingency and return contingency; candidate episodes are classified as confirmed, non-sociodual, or indeterminate before secondary coding of response orientation and substantive contribution re-formation. The framework distinguishes interaction history from current contextual inputs and separates pathway description from endpoint states and downstream outcomes. Three propositions address history-conditioned formation, pathway divergence after early episode orientations, and robustness differences among endpoint-equivalent pathways. A frozen operational protocol was calibrated on three previously unseen natural human–AI records through two separately executed model-based evaluator series, supporting operational tractability with residual disagreement concentrated at return-contingency boundaries. A supplementary exploratory analysis then compared frozen Sociodual pathway reconstructions with blind developmental/task-process segmentations and extended the comparison to three prospectively selected unseen records. Across the six examined interactions, Sociodual pathway topology was empirically non-equivalent to ordinary developmental/task topology; the distinction persisted under fine-grained re-segmentation and record-format robustness checks and replicated in all three unseen records using a fresh model-based Sociodual coding line. Socioduality therefore provides a bounded and empirically tractable process construct for analysing how human and AI contributions are formed through interaction while preserving relational pathway information that task-stage and endpoint-centred analyses do not uniquely recover.

## 1. Introduction

Artificial intelligence is increasingly used as an interactive contributor to human judgement, decision-making, problem solving, and content creation. In these settings, the analytically important object is no longer only the human, the AI, or the final product. It is also the sequence through which their contributions become connected across time.

Most human–AI evaluation still privileges capabilities, combined performance, or final outputs. Those questions remain essential, but they compress the interaction that produced the endpoint. Describing what the human produced and what the AI produced does not, by itself, constitute an analysis of human–AI interaction: interaction as a relational process becomes analytically visible only when the relation between successive contributions is preserved—when one party's response can be shown to enter the conditions under which the other party's next move is formed. Two exchanges can reach the same judgement while differing in whether the human accepted, rejected, verified, repaired, redirected, or reaffirmed in response to AI input; the same apparent human–AI combination can therefore contain materially different relational processes. Endpoint data cannot recover those histories once they have been discarded.

Empirical work already shows that sequence matters. MINT demonstrates that strong single-turn performance does not guarantee strong multi-turn performance and that models vary in their use of language feedback and tools across turns (Wang et al., 2024). Human–AI feedback-loop research shows that prior exchanges can alter later human judgements (Glickman & Sharot, 2025), while meta-analytic evidence on human–AI combinations shows that joint performance cannot be inferred mechanically from component performance alone (Vaccaro et al., 2024). These findings make the interaction process itself a legitimate explanatory target.

A growing set of relational perspectives provides important foundations. Reciprocal Human–Machine Learning examines repeated cycles of mutual learning (Te'eni et al., 2023); mutual-adaptation research formalises history-sensitive human–robot collaboration (Nikolaidis et al., 2016); coactive design centres interdependence in joint activity (Johnson et al., 2014); human–AI coevolution addresses longer-term reciprocal influence (Pedreschi et al., 2025); and Interaction-Centered Intelligence places interaction at the centre of human–AI co-creation (Davis, 2026). Together they establish a broad relational turn without fixing one common episode/pathway identification architecture.

Structural and decision traditions sharpen the comparison. IRF/IRE research has long analysed three-position exchanges (Sinclair & Coulthard, 1975; Mehan, 1979); grounding and repair research shows how understanding and interactional trouble are managed across turns (Clark & Brennan, 1991; Schegloff et al., 1977); and the Judge–Advisor System examines an initial judgement, advice, and a subsequent judgement, with Weight of Advice providing an established measure of numerical movement toward advice (Sniezek & Buckley, 1995; Bonaccio & Dalal, 2006; Bailey et al., 2023). These are substantive precedents and component overlaps, not reasons to reduce a process construct to surface sequence geometry.

Socioduality is distinguished at the level of analytical object and evidence. $A_1 \rightarrow B_1 \rightarrow A_2$ is not claimed as a new sequence form; it is the minimum candidate structure within which two contingencies are tested: $B_1$ must be formed in relation to $A_1$, and $B_1$ must then enter the observable formation of $A_2$. Confirmed episodes can overlap and link into pathways that carry interaction history forward. The construct therefore integrates sequence, reciprocity, evidence state, and pathway continuity into one bounded object rather than treating any single component as the novelty claim.

Socioduality addresses the remaining analytical gap by treating the response-conditioned formation of successive moves as an object in its own right. It refers to the unfolding relational sequence through which contributions, judgements, decisions, and actions are formed in relation to what has already occurred between two distinguishable parties. A later move may revise an earlier position, retain it after explicit consideration, reject the other party's proposal, seek verification, redirect the task, or end the interaction. What matters for construct existence is not the amount of change or the final output, but whether the required relational contingencies are observable. When confirmed episodes continue through linked moves, the pathway preserves process information that the endpoint alone cannot contain.

The article makes five contributions. First, it defines socioduality and separates construct existence from response orientation and substantive change. Second, it locates the construct within process theory, sequence organisation, feedback, grounding and repair, IRF/IRE, advice-taking, reciprocal learning, mutual adaptation, and related human–AI process research while specifying the analytical work that remains distinct. Third, it establishes nested empirical units—move, candidate/confirmed episode, sociodual pathway, and interaction container—and formalises the pathway-irreducibility principle: endpoint state does not reconstruct the relational sequence that produced it. Fourth, it develops three testable propositions concerning history-conditioned formation, pathway divergence, and robustness differences among endpoint-equivalent pathways. Fifth, it provides an identification logic that distinguishes confirmed, non-sociodual, and indeterminate evidence, separates observable termination from record cessation, and calibrates claims to observational and controlled evidentiary regimes. The central thesis is direct: the relational process through which successive human and AI moves are formed should be analysed as a primary empirical object, not recovered indirectly from final outputs. A supplementary exploratory analysis further examines the construct's analytical distinctiveness by comparing frozen Sociodual pathway reconstructions with blind developmental/task-process segmentations and extending the comparison to prospectively selected unseen records.

## 2. Theoretical Foundations and Conceptual Neighbours

### 2.1. From outcome evaluation to process explanation

Human–AI evaluation has understandably prioritised output quality, accuracy, efficiency, trust, reliance, and synergy. These questions are especially important in high-stakes settings because a combined arrangement must ultimately be judged by what it produces. Vaccaro et al. (2024), for example, provide an unusually broad assessment of whether human–AI combinations improve task performance. Their findings also point beyond simple addition: the performance of a combined system cannot be inferred mechanically from the independent performance of the human and the AI. A recent domain-specific benchmark likewise illustrates the broader output-centred orientation in human–AI evaluation by assessing general-purpose AI assistants through their visible outputs under a fixed multi-turn protocol, deliberately treating the observable output rather than hidden model processes as the analytic object (Çögenli, 2026).

The distinction developed here is not between outcomes and processes as competing objects. It is between different forms of information. Outcome analysis asks what was produced and how well it performed. Process analysis asks what happened between the parties and through what relational sequence the successive moves preceding the endpoint were formed. An endpoint state can summarise where an interaction ended, but it does not preserve the ordering, uptake, rejection, repair, verification, conflict, persistence, or role shifts through which that state emerged. An outcome may therefore be evaluated correctly while the process through which it emerged remains unknown. This problem becomes more visible as AI use shifts from one-shot recommendations toward extended, revisable, or decision-linked interaction.

The loss of this process information is theoretically consequential in its own right. Confirmed pathways remain analytically distinct even when endpoint measures are identical and even when later tests find no difference in performance or robustness. Downstream effects are therefore empirical consequences to be tested after process characterisation, not a prerequisite for the process to count as scientifically meaningful.

Process theory provides a general basis for this move. Process explanations focus on temporal sequences, events, and the mechanisms through which change and continuity unfold rather than treating phenomena solely as stable variables or end states (Van de Ven & Poole, 1995). Langley (1999) similarly emphasises that process data concern events and activities unfolding over time and require analytical strategies capable of preserving sequence. Socioduality applies this process orientation to the reciprocal, history-sensitive formation of successive moves within human–AI interaction.

The language of a history-carrying pathway should also be distinguished from organisational path dependence. In organisational theory, path dependence refers more specifically to processes in which self-reinforcing mechanisms progressively restrict alternatives and may produce lock-in (Sydow et al., 2009). Socioduality does not require self-reinforcement, narrowing choice, or lock-in. In the present framework, a sociodual pathway is the maximal uninterrupted chain of two or more overlapping confirmed episodes that preserves an empirically observed sequence of relationally conditioned moves; such a pathway may converge, diverge, repair, terminate, or reverse direction.

Several established traditions provide important foundations for this argument without constituting equivalent constructs. Conversation analysis demonstrates that interaction is sequentially organised and that subsequent turns display how prior turns have been understood and acted upon (Sacks et al., 1974). Cybernetics established the broader logic of feedback, in which prior outputs can re-enter and alter subsequent system activity (Wiener, 1948/2019). These traditions make sequence and returned influence theoretically familiar; they do not by themselves define the same human–AI process object developed here, in which observable uptake across conditionally connected moves is identified at episode level and, when linked continuation occurs, compared across pathways.

Situated-action research adds a second foundation. Suchman (1987) showed that human–machine activity cannot be adequately understood as the execution of fixed plans independent of local circumstances; action develops through the resources and contingencies of the unfolding situation. Socioduality builds on this contextual sensitivity while asking a narrower question: how does a prior human or AI response enter the observable formation of the other party's next move, and how do different patterns of uptake accumulate into distinguishable interaction pathways?

## 2.2. Multi-turn interaction and empirical feedback

Multi-turn research establishes that interactional performance cannot always be reduced to isolated prompts. MINT evaluates whether large language models use tools and natural-language feedback across repeated turns and reports that strong single-turn performance does not guarantee strong multi-turn performance (Wang et al., 2024). Ding and Tan (2026) likewise propose a structured framework for evaluating transparency, consistency, and refinement across multiple turns. These approaches show that temporally extended interaction is a distinct evaluation problem.

Feedback-loop research goes further by showing that prior human–AI exchanges may alter subsequent human cognition. Glickman and Sharot (2025) found that biased AI judgements could be internalised by human participants and amplified through repeated interaction. This evidence directly supports the idea that prior responses can be carried forward rather than disappearing after each turn.

Yet neither multiple turns nor feedback alone defines the process proposed here. A scripted dialogue may contain several turns without providing evidence that a preceding response entered the formation of the next observable move. Feedback can also be examined at aggregate, population, or system levels without identifying a two-party sequence in which responses are taken up, rejected, retained against, verified, revised, or otherwise made consequential for what follows. Socioduality therefore uses multi-turn structure and feedback sensitivity as important foundations while specifying a narrower relational process object.

### 2.3. Reciprocal learning, mutual adaptation, and coactive design

Reciprocal Human–Machine Learning is a close functional neighbour. Te'eni et al. (2023) conceptualise humans and machine-learning systems as learning partners engaged in repeated cycles of feedback. Their empirical instantiation follows domain experts and machine-learning models through several learning cycles in which each side's feedback informs subsequent classification activity. This work demonstrates that human and machine contributions may change through repeated exchange.

Socioduality is broader in one respect and narrower in another. It is broader because the next move may be relationally formed without durable learning or even without substantive change in the endpoint. A user may accept, reject, retain, verify, reformulate, resist, or tactically redirect an AI response. It is narrower because the focal object is specifically the observable sequence through which prior responses enter the formation of subsequent moves; when confirmed episodes become linked, their continuation forms a sociodual pathway. The construct does not encompass the entire architecture of joint learning.

Mutual-adaptation research in human–robot collaboration offers a related formal precedent. Nikolaidis et al. (2016) model adaptation under bounded memory, recognising that current collaborative behaviour can depend on recent interaction history. Socioduality shares this temporal sensitivity but does not require convergence, improved coordination, or successful task completion. Reciprocal updating may produce alignment, divergence, stagnation, escalation, or termination.

Coactive design also places interdependence at the centre of human–machine teamwork and seeks to support joint activity through observability, predictability, and directability (Johnson et al., 2014). Its primary contribution is normative and design-oriented: it asks how systems should be structured to support interdependent work. Socioduality is primarily analytical. It asks how one party's response enters the formation of the other party's next move, regardless of whether the interaction was well designed.

### 2.4. Interaction-centred intelligence and human–AI coevolution

Interaction-Centered Intelligence is among the closest contemporary neighbours because it explicitly proposes interaction as a primary unit of analysis. Davis (2026) emphasises interaction trajectories, coordination patterns, adaptive participation, and emergent intelligence in human–AI co-creation. This orientation strongly supports moving beyond isolated agents and outputs.

The overlap with socioduality is therefore real rather than merely thematic: both approaches reject output-only evaluation and treat interaction trajectories as analytically consequential. The distinction is that Interaction-Centered Intelligence theorises intelligence and co-creation as emergent interaction-level phenomena, whereas socioduality specifies a bounded evidentiary object with explicit episode-existence conditions, three evidence states, maximal pathway reconstruction, and an endpoint/pathway non-equivalence test. Socioduality therefore does not require creativity, emergent intelligence, successful co-creation, or adaptation; the same architecture can identify corrective, conflictual, routine, resistant, or deteriorating exchanges.

Human–AI coevolution addresses a broader temporal and systemic scale. Pedreschi et al. (2025) define coevolution as a continuing process in which humans and AI algorithms influence one another's development. Socioduality may occur within such systems, and repeated sociodual processes may contribute to longer-term coevolution. Nevertheless, coevolution should not be reduced to bounded dyadic sequences, and a sociodual sequence need not produce durable evolutionary change.

### 2.5. Structural precedents, grounding, repair, and advice-taking

The $A_1 \rightarrow B_1 \rightarrow A_2$ form has clear structural precedents and is not claimed as a novel sequence geometry. Sinclair and Coulthard's (1975) classroom discourse model and Mehan's (1979) related work describe recurrent three-move Initiation–Response–Feedback/Evaluation exchanges in which a teacher initiates, a learner responds, and the teacher follows up or evaluates. This literature is important because it shows that a three-position exchange can be a stable analytical unit. Its focal object, however, is the organisation of classroom discourse and institutionally differentiated teacher–student roles. In socioduality, $A_1 \rightarrow B_1 \rightarrow A_2$ serves only as a minimum evidentiary configuration; $A_2$ may be a judgement, rejection, reaffirmation, verification request, action, or termination, and the analytical object is the history-carrying relational episode and its possible linkage into a pathway rather than a classroom exchange grammar.

Grounding, repair, and conversation analysis provide particularly close technical foundations for the evidentiary logic. Clark and Brennan (1991) describe communication as coordinated activity that depends on common ground and its moment-by-moment accumulation, while repair research identifies organised practices for dealing with problems in speaking, hearing, and understanding (Schegloff et al., 1977). More specifically, conditional relevance captures how a recognisable first pair part makes a fitted second pair part relevant (Schegloff, 2007), and the next-turn proof procedure treats a subsequent turn as endogenous evidence of how a prior turn was understood (Sacks et al., 1974). Socioduality does not claim these principles as novel. It extends their sequential evidentiary insight into a bounded human–AI identification architecture in which candidate episodes receive explicit confirmed, non-sociodual, or indeterminate status, construct existence is separated from response orientation and substantive re-formation, and linked confirmed episodes are reconstructed as maximal pathways that can be compared even when endpoint states coincide.

Advice-taking research is a particularly close empirical neighbour for human–AI judgement settings. The Judge–Advisor System studies how a decision-maker responds to advice, often by comparing an initial judgement with a final judgement after advice (Sniezek & Buckley, 1995; Bonaccio & Dalal, 2006). In numerical tasks, Weight of Advice quantifies the degree to which the final estimate moves toward the advisor's estimate and has become a widely used operational measure (Bailey et al., 2023). This literature provides a mature way to measure one important response orientation. It does not, however, require reciprocal conditioning of the advisor's contribution by the judge's prior move, nor does recursive, history-carrying continuation constitute the defining analytical object across JAS designs. Socioduality should therefore complement rather than relabel advice-taking: WOA can quantify directional adjustment within some confirmed sociodual episodes or pathways, while episode/pathway analysis retains response orientation, prior history, continuation, repair, and other process information that WOA alone does not encode.

### 2.6. Locating socioduality within prior theory

Construct development requires a specified analytical object, clear boundaries, relations to prior theory, and empirical implications (Suddaby, 2010). Socioduality organises established insights about sequence, feedback, reciprocity, situated action, and process around one bounded object: the observable, history-sensitive relational sequence through which successive human and AI moves are formed in relation to prior exchange and represented empirically as confirmed episodes and linked pathways.

Prior literature can relate to this construct in different ways. Some traditions provide intellectual foundations, such as sequence organisation, feedback, situated action, and process theory. Other studies contain component-level overlaps because they observe revision, reciprocal influence, uptake, or changing contributions while explaining a different focal phenomenon. Still others define domain-specific processes, such as reciprocal learning or mutual adaptation, that may instantiate particular pathway patterns. These relationships should not be treated as equivalent forms of conceptual competition.

Section 5 applies this redundancy test explicitly by asking whether established theoretical and methodological components already provide the same bounded analytical object and identification architecture.

## 3. Defining Socioduality

### 3.1. Rationale for the term

The term socioduality combines socio-, indicating relational production, with duality, indicating two analytically distinguishable poles within the focal process. The prefix does not restrict the construct to conventional social relationships between human individuals. It indicates that the developing moves are produced relationally rather than being treated as fixed contributions that exist independently of the interaction. Duality does not imply equal power, identical capacities, metaphysical symmetry, or moral equivalence. It refers to the presence of two distinguishable parties whose successive moves are conditionally connected.

The use of duality also requires distinction from Giddens's (1984) "duality of structure." In structuration theory, duality refers to structure as both medium and outcome of recursively organised social practices. Socioduality does not adopt that meaning or claim continuity with structuration theory. Here, duality denotes the two analytically distinguishable poles whose successive moves are conditionally connected within the focal interaction. The terminological overlap is therefore acknowledged explicitly while the theoretical objects remain different. The label is likewise not used as a synonym for macrostructural uses of "social dualism"; the present construct refers to a bounded sequential relational process.

A party may be an individual, a group, a team, a board, an organisation, a role, a technological system, or another analytically distinguishable decision unit. Treating an AI system as a party is an analytical decision, not an attribution of consciousness, legal personhood, or moral responsibility. The concept identifies the position from which an observable move is generated within the sequence.

### 3.2. Definition

Socioduality is a sequential, reciprocal, and history-carrying relational process between two distinguishable parties in which a response from one party becomes part of the observable conditions under which the other party's subsequent contribution, judgement, decision, or action is formed. The analytical object is the conditionally connected relational sequence—represented empirically as a confirmed episode and, when linked continuation is established, as a sociodual pathway—rather than the endpoint alone.

The definition contains four linked elements. First, the parties must be distinguishable. Second, the process must unfold sequentially. Third, the relation must be reciprocal in the minimal sense that an observable move by one party conditions a response from the other and that response, in turn, enters the observable formation of the first party's subsequent move. Fourth, the process is history-carrying: once an exchange has occurred, subsequent moves arise under conditions that include the available interaction history as well as any new contextual inputs. At the minimum episode level, history-carrying is instantiated in the limited sense that $B_1$ becomes part of the conditions under which $A_2$ is formed; in longer pathways, earlier accumulated history may also remain consequential beyond the immediately preceding response. Substantive change is not required. A later move may adopt, reject, retain, revise, verify, redirect, or terminate in relation to the preceding response. Material or substantive contribution re-formation is therefore treated as an important process property, not as the constitutive definition of socioduality.

### 3.3. Unit of analysis

The empirical architecture is nested rather than singular. Socioduality should not be assigned automatically to an entire conversation, session, or task merely because one reciprocal episode occurs within it. Four levels should be distinguished:

- Move: one observable contribution, response, judgement, decision, action, or explicitly recorded termination produced by one party.
- Sociodual episode: the minimum confirmed $A_1 \rightarrow B_1 \rightarrow A_2$ configuration in which $B_1$ is responsive to $A_1$ and there is sufficient evidence that $B_1$ entered the formation of $A_2$.
- Sociodual pathway: the maximal uninterrupted chain of two or more overlapping confirmed episodes, for example $A_1 \rightarrow B_1 \rightarrow A_2$ followed by $B_1 \rightarrow A_2 \rightarrow B_2$.
- Interaction/session container: the broader recorded conversation, task, meeting, or workflow within which zero, one, or multiple sociodual episodes or pathways may occur.

This nesting prevents episode-level evidence from being inflated into a session-level label. If only one three-move segment in a long interaction satisfies the evidentiary criteria, it is coded as a confirmed sociodual episode, not as a sociodual pathway. A pathway is established only when at least one further overlapping candidate episode is confirmed; indeterminate or non-sociodual candidates break the chain and are not bridged by topic continuity or background history. Pathway length, density, continuity, and interruption can then be studied as properties rather than assumed from the existence of a session.

The analytical object remains the relational sequence—represented as a confirmed episode or, when linked episodes are present, a sociodual pathway—rather than either party or the endpoint state alone. Endpoint state refers here to the terminal observable state of the focal task, episode, or bounded pathway, such as a final judgement. Downstream outcome refers to later variables assessed after process characterisation, such as accuracy, robustness, reproducibility, burden, or oversight. Keeping these terms separate prevents the word outcome from conflating where a process ended with what consequences followed from it.

### 3.4. Necessary conditions

Three conditions are necessary for confirming a sociodual episode:

- Two distinguishable parties: the researcher can identify the two poles producing the focal moves.
- Response contingency: $B_1$ is produced in relation to $A_1$ rather than independently, merely in parallel, or according to a fixed response that $A_1$ cannot alter.
- Return contingency: the available evidence supports the claim that $B_1$ entered the formation of $A_2$—that $A_2$ presupposed or advanced a state established or altered by $B_1$. Continuation of a broader task topic, temporal proximity, or a fresh initiation prompted by new contextual input does not by itself satisfy this requirement.

These conditions establish whether an episode is confirmed; they should be assessed before coding what kind of response occurred. Sequentiality is instantiated by the ordered $A_1 \rightarrow B_1 \rightarrow A_2$ candidate structure, and minimum history-carrying by the return-contingency requirement; they therefore do not operate as additional independent episode-confirmation gates. Labels such as acceptance, rejection, reaffirmation, revision, verification-seeking, redirection, or observable termination are response orientations within an already confirmed episode. They do not, by themselves, prove that relational contingency existed. Substantive change in content, criteria, strategy, judgement, or action is likewise not required. When an outcome-relevant change is observable, it can be coded separately as substantive contribution re-formation. Scripted or rule-based interaction is not excluded merely because it is rule-based, but a fixed alternation that cannot be altered by the preceding focal move fails the contingency test. In sustained human–AI dialogue, response contingency may often be readily supported because the AI response is generated in relation to the immediately preceding human move; the more discriminating evidentiary burden may therefore fall on return contingency. Retaining both requirements preserves the reciprocal architecture and permits either direction to be coded as absent or insufficient where the record warrants it.

### 3.5. Minimum sociodual episode and pathway extension

The minimum sequence capable of confirming a sociodual episode is:

$$A_1 \rightarrow B_1 \rightarrow A_2$$

The three-position form is an established structural precedent rather than the contribution itself. Here, $A_1 \rightarrow B_1 \rightarrow A_2$ is the minimum evidentiary configuration for testing response and return contingency. $A_2$ may be a judgement, decision, acceptance, rejection, reaffirmation, revision, verification request, observable action, or observable termination. If $B_2$ is then formed in relation to $A_2$, the overlapping $B_1 \rightarrow A_2 \rightarrow B_2$ sequence becomes a further candidate episode; when confirmed, linked episodes establish a sociodual pathway. Longer pathways are therefore built from connected evidence-bearing episodes rather than by labelling an entire session sociodual by default.

### 3.6. Episode confirmation versus response orientation

Temporal succession is not equivalent to episode confirmation. The primary identification question is whether the two relational links required for the episode are supported: $B_1$ must be responsive to $A_1$, and $B_1$ must enter the observable formation of $A_2$. Only after this evidentiary decision should $A_2$ be coded for response orientation. Acceptance, rejection, reaffirmation, revision, verification-seeking, redirection, or observable termination describe what occurred within the confirmed episode; they are not exhaustive proof that an episode exists. An unchanged endpoint state is therefore compatible with a confirmed episode when the record explicitly shows that $B_1$ was considered and rejected or used in reaffirmation. By contrast, an identical $A_1$ and $A_2$ with no evidence linking $B_1$ to $A_2$ remains indeterminate in passive data. Controlled designs can manipulate $B_1$, preserve pre/post judgements, elicit reasons or verification behaviour, and collect supplementary traces to test causal sensitivity without making hidden cognition part of the construct.

### 3.7. Exclusion and indeterminate cases

Not every recorded two-party sequence qualifies as a confirmed sociodual episode. Clear non-cases include one-way transmission with no subsequent observable move by the receiving party; independently generated human and AI outputs later combined by a researcher or third party; $B_1$ outputs generated independently of $A_1$; and $A_2$ moves known by design to be precommitted or generated independently of $B_1$. A fixed or scripted alternation is non-sociodual when the preceding focal move cannot alter the formation of the next move. A record containing $A_1$, $B_1$, and $A_2$ timestamps is therefore insufficient unless both relational contingencies are supported.

A second category is indeterminate evidence rather than a positive or negative case. In passive logs, for example, an unchanged $A_2 = A_1$ with no explicit reference, revision trace, verification behaviour, or other evidence may not reveal whether $B_1$ was considered, ignored, or merely seen. Such cases should be coded as indeterminate and excluded from confirmed pathway comparisons unless a stronger evidentiary regime resolves the ambiguity. Indeterminacy is a property of the evidence, not proof that the process is absent or present. Mere record cessation after $B_1$ is also not an $A_2$ termination move. Termination counts as $A_2$ only when an observable decision or action to terminate is recorded and its relation to $B_1$ is supported. Likewise, the absence of a further linked confirmed episode does not by itself establish termination; a confirmed episode with no further linked episode in the available record is simply unextended in that record unless an observable termination move is present. Finally, confirmation of one episode within a long session does not confirm the whole session; coding remains bounded to the episode or to linked episodes that form a pathway.

### 3.8. Scope conditions

Socioduality does not require equal influence, successful coordination, conscious intention, durable learning, substantive revision, a positive endpoint state, or continuous interaction. A party may accept, reject, retain, verify, misinterpret, resist, redirect, or amplify the other party's response. The process may be asymmetric, conflictual, interrupted, highly constrained, observably terminated, or harmful. External information, institutional constraints, new documents, changed objectives, and other current conditions may also enter the sequence; socioduality is relationally dyadic but contextually open. External contextual inputs may affect a dyadic episode without becoming a third focal party, but an independently acting third party whose contribution itself functions as a focal interactional move makes that segment triadic rather than merely contextually open. The present article specifies and evaluates the construct only for human–AI dyads. AI–AI and other non-human dyads are outside the present claim rather than automatically included or excluded by the generic party definition. Likewise, genuinely triadic or network interaction is outside the current analytical architecture. Scope eligibility should therefore be assessed before evidentiary classification: structures outside the present human–AI dyadic scope are not classified as confirmed, non-sociodual, or indeterminate under this specification. If a team, board, or organisation is treated as one party in later applications, its internal multi-actor dynamics must be analytically bracketed rather than silently collapsed into the dyadic process. Broader portability requires separate theorisation and evidence.

## 4. The Socioduality Process Model

### 4.1. Observable sequence

As shown in Figure 1, the model represents a sequence of moves from which sociodual episodes and pathways may be identified. $A_1 \rightarrow B_1 \rightarrow A_2$ is the minimum confirmed episode when both relational contingencies are supported. If $B_2$ is responsive to $A_2$, $B_1 \rightarrow A_2 \rightarrow B_2$ becomes a candidate overlapping episode; when that further episode is confirmed, the two linked episodes establish a sociodual pathway. The arrows therefore indicate empirically supported relational conditioning, not chronology alone.

**Figure 1. The Socioduality Process Model**

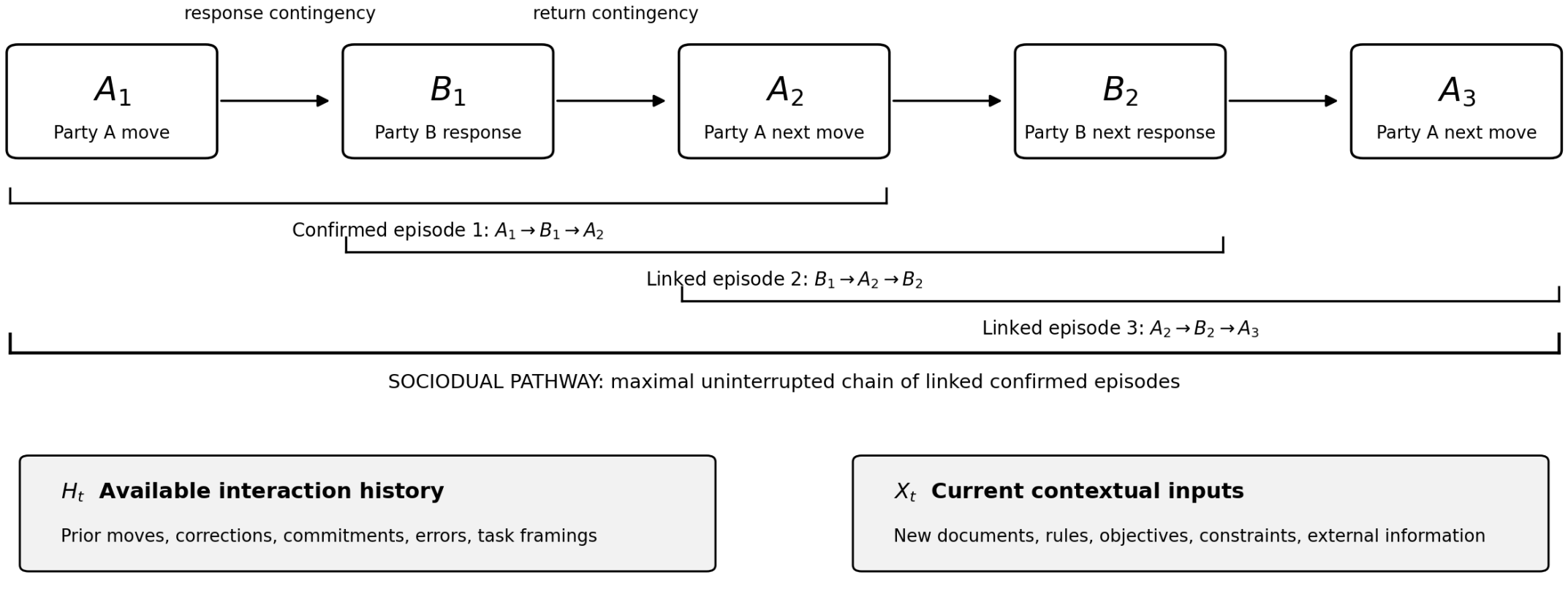

Note. Arrows represent relational conditioning to be established empirically, not chronology alone. A confirmed episode requires both response contingency ($A_1 \rightarrow B_1$) and return contingency ($B_1 \rightarrow A_2$). Candidate sequences with insufficient evidence remain indeterminate and do not extend a pathway. Two or more overlapping confirmed episodes form a maximal uninterrupted sociodual pathway. $H_t$ denotes the available interaction history and $X_t$ current contextual inputs; endpoint states and downstream outcomes are analytically separate and cannot reconstruct the pathway.

In human–AI interaction, the minimum episode can be written as human move$_1$ → AI response$_1$ → human move$_2$. The second human move is analytically important even when its endpoint value is unchanged, but its orientation should be coded only after the episode-level contingency has been established. If a substantive, outcome-relevant alteration relative to the same party's preceding observable contribution or state occurs in content, criteria, strategy, judgement, or action, it can be coded separately as contribution re-formation. If the interaction continues and a further linked episode is confirmed, the two episodes establish a sociodual pathway.

### 4.2. Interaction history and current conditions

The observable chain shows order but does not imply that each move is produced only by the immediately preceding response. For analysis, two additional sources should be distinguished: the accumulated interaction history available before a move and current conditions or new inputs introduced at that point. These distinctions are analytical aids rather than a predictive mathematical model.

Interaction history may include earlier claims, corrections, commitments, errors, and task framings that remain consequential for later contributions. Current conditions may include a newly supplied document, changed objective, time constraint, institutional rule, or other input that was not generated by the preceding exchange. Separating these sources prevents every later change from being attributed automatically to the other party's response.

### 4.3. History-carrying but contextually open interaction

Interaction history is structured rather than homogeneous. Recent responses may remain especially visible, while earlier elements may persist, weaken, be reinterpreted, or disappear. Socioduality therefore assumes history sensitivity, not perfect memory or equal weighting of all prior moves.

New contextual conditions can redirect a pathway without becoming a third focal party. Human–AI socioduality is therefore analytically dyadic but contextually open: the two focal parties remain the source of the conditionally connected moves while documents, rules, objectives, interfaces, system constraints, or external events may alter the conditions under which those moves are formed.

### 4.4. Non-determinism and relational conditioning

The model does not imply that the same $B_1$ must always produce the same $A_2$. Parties may interpret the same response differently, attend to different elements, or face different current conditions. Socioduality therefore proposes relational conditioning rather than mechanical determination: prior exchange changes the conditions of subsequent contribution formation without uniquely determining what the next contribution must be.

### 4.5. An open, history-carrying sequence

The process should not be represented as a closed circle returning to an unchanged starting point. $A_2$ does not arise under the same conditions as $A_1$, and $B_2$ does not respond to the same input as $B_1$. Each move changes the available interaction history that may become consequential for subsequent moves.

> Socioduality does not treat interaction as a neutral channel connecting independent contributions. Interaction is the relational process through which successive contributions, judgements, decisions, and actions are formed in relation to what has already occurred.

## 5. Conceptual Positioning and the Integrated Redundancy Test

The preceding section establishes the main intellectual foundations, structural precedents, and direct neighbours. Table 1 summarises their conceptual status without repeating the full literature review. The central redundancy question is whether these traditions, separately or in an established integration, already provide the same bounded analytical object, episode/pathway units, evidentiary rules, and empirical programme.

### 5.1. Why not process theory + JAS + sequential interaction analysis?

A researcher can combine process-theoretic sequence preservation, sequential interaction coding, and JAS/advice-taking measures to analyse many of the same human–AI records. Socioduality's construct claim is not that these component methods are unavailable elsewhere. It is that they do not, by themselves, establish one common process object with explicit episode/pathway units, a calibrated evidentiary architecture for contingency assessment, and a research programme centred on reciprocal response-conditioned pathway formation.

This claim is falsifiable rather than rhetorical. An established integration that captures the same object and identification work with equal or greater parsimony would render a separate construct redundant. On the present specification, however, socioduality adds a stable analytical architecture: it fixes what counts as a candidate episode, what evidence confirms or rejects relational contingency, when uncertainty remains indeterminate, how confirmed episodes link into pathways, and why endpoint and downstream outcomes must remain analytically separate. Its distinctiveness should therefore be judged by comparative empirical use, not by label novelty.

**Table 1. Conceptual distinctions**

| Literature / concept | Status in relation to socioduality | Central analytical concern | Relationship to socioduality |
|---|---|---|---|
| Conversation analysis / sequence organisation and repair | Intellectual foundation / component overlap | Sequential organisation of social action, displayed understanding, and repair of interactional trouble | Provides the closest technical precedent for response/return evidentiary logic through conditional relevance and the next-turn proof procedure. Socioduality does not claim these principles as novel; it adds explicit episode evidence states, maximal pathway reconstruction, and endpoint/pathway comparison within a bounded human–AI process object. |
| Grounding / common ground | Intellectual foundation / component overlap | Moment-by-moment coordination of shared knowledge, assumptions, and mutual understanding | Supports history-carrying coordination; socioduality does not require successful grounding or shared understanding and also includes rejection, failed grounding, conflict, error propagation, and termination. |
| IRE/IRF three-move exchange | Structural precedent / component overlap | Initiation–Response–Evaluation/Feedback patterns in classroom discourse | Shows that three-move interaction units are established and not novel. Socioduality uses $A_1 \rightarrow B_1 \rightarrow A_2$ as a minimum evidentiary form for a response-conditioned episode that may link into a pathway, not as a classroom-role exchange grammar. |
| Judge–Advisor / advice-taking | Direct empirical neighbour / measurement precedent | How a decision-maker uses advice; often measured as movement from an initial to a final judgement | Directly overlaps with terminal $H_1 \rightarrow$ advice $\rightarrow H_2$ cases and offers WOA as a useful measure. Reciprocal conditioning of both parties and recursive history-carrying pathway continuation are not defining requirements of the JAS paradigm. |

| **Literature / concept** | **Status in relation to sociodualitiy** | **Central analytical concern** | **Relationship to socioduality** |
|---|---|---|---|
| Cybernetics / feedback | Intellectual foundation / component overlap | Returned influence, feedback, regulation, and changing system states | Provides the general logic of feedback; socioduality specifies an analytically dyadic, history-sensitive sequence in which uptake can take the form of acceptance, rejection, retention, revision, verification, or redirection. |
| Situated action | Intellectual foundation | Action as locally produced through unfolding circumstances rather than fixed plans alone | Supports contextual and non-deterministic process formation; it does not itself define the bounded sociodual episode/pathway architecture as the focal analytical object. |
| Multi-turn interaction | Empirical setting / structural overlap | More than one exchange | Creates a temporal setting in which socioduality may be observed; multiple turns alone are insufficient without evidence of relational uptake. |
| Reciprocal learning | Direct process neighbour / domain-specific process | Mutual knowledge or capability change | Shares iterative mutual influence but requires learning; socioduality also includes rejection, retention, temporary uptake, verification, termination, and other non-learning process forms. |
| Mutual adaptation | Direct process neighbour / domain-specific process | Behavioural or strategic adjustment | Shares history-sensitive interaction; adaptation is one possible pathway pattern rather than a defining condition of socioduality. |
| Coactive design | Adjacent normative framework | Designing support for interdependent joint activity | Prescribes conditions for effective interdependence rather than analysing how observable moves are formed across bounded episodes and pathways. |
| Interaction-Centered Intelligence | Direct conceptual neighbour | Interaction trajectories, coordination, adaptive participation, and emergent intelligence | Shares interaction as a primary unit and emphasises trajectories and coordination. Socioduality adds explicit episode-existence conditions, confirmed/non-sociodual/indeterminate evidence states, maximal pathway reconstruction, and endpoint/pathway non-equivalence without requiring emergent intelligence, creativity, successful co-creation, or adaptation. |

| Literature / concept | Status in relation to socioduality | Central analytical concern | Relationship to socioduality |
|---|---|---|---|
| Human–AI coevolution | Adjacent macro-process | Long-term systemic mutual influence | Operates at a broader temporal and systemic scale; bounded sociodual pathways may contribute to but are not equivalent to coevolution. |
| Performance and synergy | Complementary outcome-level literature | Quality of human, AI, and combined outcomes | Provides endpoint measures that can be compared after pathways are characterised; it does not by itself reveal how the endpoint was relationally formed. |
| Organisational path dependence | Related process tradition / terminology requiring differentiation | Self-reinforcing mechanisms, narrowing alternatives, and potential lock-in over time | Shares the premise that history can matter, but socioduality does not require self-reinforcement, restricted alternatives, or lock-in; "pathway" here denotes two or more linked confirmed episodes that preserve a relational sequence rather than organisational path dependence. |

Shared ancestry is therefore a foundation rather than a disqualification. The added value of socioduality lies in making relational process variation consistently identifiable and comparable through a single episode/pathway architecture that can be tested against alternative analytical integrations.

## 6. Illustrative Applications

Human–AI interaction is the focal domain of the present construct because prompts, responses, judgements, revisions, decisions, and contextual additions can often be preserved in chronological order. The example below is illustrative rather than confirmatory and is intended to show the analytical object without implying that every multi-turn exchange is equally informative.

### 6.1. Human–AI judgement

Consider a user or professional making an initial judgement $H_1$ and then presenting the case to a generative AI system. The AI produces $AI_1$ in relation to that initial move. $H_2$ is the human's next observable move after encountering $AI_1$. $H_2$ may accept the recommendation, reject it, retain the original judgement while explicitly responding to the AI's reasons, revise the decision, seek verification, redirect the task, take an observable action, or terminate the interaction. $H_2$ is therefore analytically important even when $H_2$ and $H_1$ share the same endpoint value. If the interaction continues, $H_2$ may then become part of the conditions under which $AI_2$ is produced.

In numerical advice-taking tasks, established JAS measures can be retained within this analysis. WOA can quantify how far $H_2$ moves from $H_1$ toward $AI_1$ and therefore provides useful information about directional advice utilisation. Socioduality does not replace that measure. It adds information that WOA alone does not necessarily preserve: for example, an unchanged numerical $H_2$ can yield WOA = 0 whether the record shows explicit consideration and rejection of $AI_1$ or provides no observable uptake at all. The first case is a confirmed response orientation; the second may remain indeterminate in passive data. Combining established advice-taking metrics with episode/pathway coding is therefore preferable to treating them as competing approaches.

A correct or unchanged final judgement can conceal sharply different confirmed pathways: AI correction of a human omission, human correction of an AI error, explicit rejection followed by reaffirmation, verification before acceptance, progressive mutual refinement, or late repair after error propagation. These are not interchangeable process records simply because their endpoints coincide. Socioduality preserves the interactional history, response orientations, repair and verification events, and asymmetries through which the endpoint was reached, and then permits researchers to test what those pathway differences subsequently predict.

### 6.2. Boundary illustrations: non-case and indeterminate case

The construct boundaries can be illustrated briefly. If a human and an AI independently evaluate the same case and their outputs are combined only afterward, no reciprocal episode exists. If $H_1 \rightarrow AI_1 \rightarrow H_2$ is observed but the available record does not show whether $AI_1$ entered the formation of $H_2$, the candidate remains indeterminate. Classification also remains local within longer sessions: a single confirmed episode does not establish a session-wide Sociodual pathway.

## 7. Theoretical Propositions

The propositions translate the construct into three testable consequences. Proposition 1 addresses history-conditioned formation beyond the minimum history-carrying required for episode confirmation. Proposition 2 links early confirmed response orientations to subsequent pathway development. Proposition 3 tests whether specified differences in pathway composition predict robustness among endpoint-equivalent cases. All three build on the pathway-irreducibility principle: endpoint information does not reconstruct the relational sequence that produced it.

Pathway-irreducibility principle. An endpoint state identifies where a bounded process ended but cannot reconstruct the sequence of response orientations, corrections, rejections, verifications, repairs, continuations, or other relational events through which that state was produced. Confirmed sociodual episodes and pathways are therefore empirical objects in their own right; downstream consequences are a subsequent question, not the source of their existence.

### 7.1. History-conditioned subsequent formation

Within sociodual pathways, accumulated interaction history may contribute information about the next observable move beyond the immediately preceding response and current conditions. This stronger history-sensitivity implication can be tested by holding the focal response and current task conditions constant or closely matched while varying the accessible prior history—for example, histories of accurate versus inaccurate AI advice, prior correction, repeated disagreement, verification, or successful versus failed repair. The prediction concerns systematic differences in observable response orientation or subsequent moves, not an assumption that every individual will react differently.

> Proposition 1. Within sociodual pathways, holding the focal response and current conditions constant or closely matched, differences in accessible prior interaction history will be associated with systematic differences in the distribution of subsequent observable response orientations or moves.

### 7.2. Early episode orientation and subsequent pathway divergence

Once an episode is confirmed, different early response orientations may alter whether linked continuation occurs and, where it does, what kinds of later configurations become possible or likely. To avoid a definitional result, this claim concerns early episodes that do not themselves observably terminate the interaction. For example, verification-seeking may open a sequence of evidence checks, whereas rejection, reaffirmation, or redirection may produce different continuation structures even when starting conditions and immediate endpoint values are comparable. This is a process-to-process claim: it concerns how early episode organisation relates to subsequent continuation and pathway development, not whether one pathway is better.

> Proposition 2. Under comparable starting human–AI–task configurations, and among early confirmed episodes that do not themselves observably terminate the interaction, differences in response orientation will be associated with systematic differences in subsequent continuation structure, including whether a linked sociodual pathway forms and, if so, its verification, repair, redirection, reaffirmation, or other configuration.

### 7.3. Endpoint-state equivalence and hidden robustness

A downstream implication arises when endpoint states are identical. The relevant predictor is not pathway difference in the abstract but pathway composition: whether the endpoint was reached through confirmed verification, correction, repair, reaffirmation, rejection, unexamined acceptance, compensating error, or combinations of these episode orientations and re-formation events. Equivalent endpoint states do not establish equivalent pathways; the additional empirical question is whether these specified configurations predict robustness when the task is repeated, perturbed, or exposed to new error.

> Proposition 3. Among sociodual pathways with equivalent endpoint performance, differences in pathway composition—particularly the presence and configuration of confirmed verification, correction, repair, unexamined acceptance, and compensating-error episodes—will predict differences in downstream robustness under repetition, perturbation, or error introduction.

## 8. Studying Socioduality Empirically

### 8.1. Nested analytical units and data structure

Empirical research should preserve time order while distinguishing the coding unit from the data container. Scope eligibility is assessed first: the present specification applies to focal human–AI dyads, while genuinely triadic or network segments remain outside the current classification architecture. Within an eligible dyad, the minimum coding unit is the candidate three-move episode $A_1 \rightarrow B_1 \rightarrow A_2$. It is confirmed only when $B_1$ is responsive to $A_1$ and the available evidence supports the claim that $B_1$ entered the formation of $A_2$. Two or more overlapping confirmed episodes form a maximal uninterrupted pathway; a longer conversation, task, or session is the container and may contain zero, one, or multiple confirmed pathways. Suitable data may include conversation records, prompt–output logs, document revisions, time-stamped judgements or decisions, action records, negotiation transcripts, and controlled experimental records. Reporting should state the number of candidate episodes, confirmed episodes, non-sociodual episodes, indeterminate episodes, out-of-scope segments where relevant, and how confirmed episodes were linked into pathways.

Retrospective statements such as "the AI influenced me" may be useful but are insufficient on their own. The strongest evidence allows researchers to compare what preceded the response, what the response introduced, and how the following move related to it. Full access to internal cognition is neither required nor presumed. Claims should remain at the level supported by observable traces: an internal belief change should not be inferred unless it is expressed in a relevant observable move.

### 8.2. Episode identification and response orientation

Episode identification and response-orientation coding should occur in that order. First, the researcher evaluates whether the required response and return contingencies are sufficiently supported to classify an in-scope candidate episode as confirmed, non-sociodual, or indeterminate. Only confirmed episodes are then coded for orientation such as acceptance/adoption, rejection, reaffirmation, revision, verification-seeking, redirection, or observable termination. Response-orientation codes are not necessarily mutually exclusive when the record supports more than one function; for example, rejection may co-occur with reaffirmation, verification with adoption, or rejection with observable termination. These categories describe process content; they do not create evidence of contingency merely because an $A_2$ action can be assigned a label. Continuation status should be coded separately from response orientation. A confirmed episode is extended only when a further linked confirmed episode is established. If no further linked episode is observed, the episode is coded as unextended in the available record, not as terminated. Observable termination is reserved for a recorded termination decision or action whose relation to the preceding response is supported. Established advice-taking measures such as WOA can complement post-confirmation analysis but should not be used as episode-identification rules.

### 8.3. Evidentiary regimes: observational and experimental

Claims about relational contingency should be calibrated to the evidentiary regime rather than treated as equally observable in all datasets. In an observational or log regime, researchers rely on naturally occurring traces such as explicit references, revisions, response orientations, source use, and temporal records. These data can strongly support relational links when the relation is visible, but silent cases may remain indeterminate. In a controlled experimental regime, researchers can randomise exposure to $B_1$ or manipulate its content, correctness, framing, or strength while preserving comparable starting conditions; they can also collect pre/post judgements, elicited justifications, verification choices, or other observable actions. Such designs can test causal sensitivity to $B_1$ without claiming direct observation of internal cognition. Additional behavioural or telemetry channels may strengthen the evidentiary record, but their role should be specified explicitly and claims should remain limited to what those channels make observable.

For record-based coding, “observable” is relative to the assigned data channel rather than to everything that may have occurred outside it. Future studies may enlarge the evidentiary channel through screen recording, video, direct behavioural observation, interface logs, think-aloud protocols, or post-task self-report. These channels should retain their distinct epistemic status: an explicit move in a transcript, an observed behavioural termination, and a later self-report about why the interaction ended are different forms of evidence and should not be collapsed into one another.

### 8.4. Substantive contribution re-formation and causal evidence

Material or substantive contribution re-formation should be coded as a process property rather than a condition for identifying socioduality. It occurs when a prior response is associated with an observable, outcome-relevant alteration in the same party's subsequent contribution or observable state relative to that party's preceding contribution or state, including content, problem framing, decision criteria, evidence use, option set, reasoning strategy, recommendation, judgement, or action. Mere consideration, acceptance, rejection, reaffirmation, verification, or termination does not by itself establish substantive re-formation; any of these orientations may co-occur with re-formation only when an outcome-relevant alteration is separately observable. Lexical polishing, formatting, or paraphrase without functional change should not be coded as substantive re-formation. Small changes can nevertheless be material when they alter an outcome-relevant parameter. Causal attribution should be calibrated to the design: controlled experiments can manipulate $B_1$ while preserving comparable conditions, whereas observational data rely more heavily on explicit references, source tracing, revision histories, matched sequences, or other naturally occurring evidence. The counterfactual question—whether a meaningfully different $B_1$ would have altered $A_2$ or another observable response orientation—is therefore a stronger causal test, not a definitional threshold.

### 8.5. History and current conditions

Researchers should separate prior interaction history from new contextual inputs. A later move may be formed in relation to the other party's response, a newly supplied document, a changed objective, or a combination of these influences. Recording current contextual inputs prevents every later move from being attributed automatically to the dyadic exchange. Longer sequences can also reveal whether earlier elements persist, weaken, reappear, or alter how the most recent response is interpreted or taken up.

### 8.6. Process, endpoint state, and downstream outcomes

Sociodual analysis begins with confirmed episodes and pathways rather than using an endpoint state to infer the process retrospectively. The endpoint state is the terminal observable state of the focal task, confirmed episode, or bounded pathway, such as a final judgement; downstream outcomes are later variables such as accuracy under repetition, robustness, reproducibility, oversight, burden, or governance consequences. The first empirical product is therefore episode and pathway characterisation itself. Researchers may then test downstream associations as a second analytical layer. Failure to find such associations would not erase confirmed process differences; it would show that those differences were not consequential for the particular downstream outcome under the studied conditions. The empirical identification architecture is summarised in Table 2.

**Table 2. Initial empirical identification framework**

| Analytical task | Guiding question | Possible evidence |
|---|---|---|
| Establish scope and data container | Is the focal structure within the present human–AI dyadic scope, and what larger conversation, task, session, or workflow is being sampled? | Identify the focal human–AI dyad and recorded container boundaries. Genuinely triadic or network segments are outside current evidentiary classification rather than non-sociodual or indeterminate. |
| Identify candidate episodes | Where do alternating three-move sequences $A_1 \rightarrow B_1 \rightarrow A_2$ occur? | Time-stamped logs, transcripts, judgement records, action records, revision histories. |
| Test response contingency | Was $B_1$ formed in relation to $A_1$ rather than independently or by an invariant script? | Content-contingent response, experimental design, system trace, or other evidence linking $A_1$ to $B_1$. |
| Test return contingency | Did $B_1$ enter the observable formation of $A_2$? | Explicit reference, designed manipulation, revision/source trace, justification, verification behaviour, or other evidence appropriate to the data regime. |
| Assign evidentiary status | Is the candidate episode confirmed, non-sociodual, or indeterminate? | Confirmed when both contingencies are supported; non-sociodual when a required contingency is shown absent; indeterminate when evidence is insufficient. |
| Code response orientation | Within confirmed episodes, what did $A_2$ do in relation to $B_1$? | Acceptance/adoption, rejection, reaffirmation, revision, verification-seeking, redirection, observable termination, or domain-specific orientation. |
| Code post-confirmation properties | Did substantive re-formation occur, and what is the continuation status of the confirmed episode? | Outcome-relevant alteration relative to the same party's preceding observable contribution or state; separately code unextended in the available record, extended through a further linked confirmed episode, or observable termination. |

| Analytical task | Guiding question | Possible evidence |
|---|---|---|
| Link episodes into pathways | Which maximal uninterrupted chains of two or more overlapping confirmed episodes form pathways? | Overlapping episodes such as $A_1$–$B_1$–$A_2$ and $B_1$–$A_2$–$B_2$; indeterminate or non-sociodual candidates break the chain and are not bridged by topic continuity or background history. A single confirmed episode is not by itself a sociodual pathway. |
| Separate current contextual inputs | What new information or conditions entered outside the dyadic history? | New document, rule, objective, constraint, interface condition, third-party information, or contextual event. |
| Separate endpoint state and downstream outcomes | Where did the focal process end, and what later consequences are tested? | Endpoint state such as final judgement; downstream outcomes such as robustness, repeated-task accuracy, reproducibility, oversight, burden, or governance consequences. |

The empirical programme is deliberately staged rather than reduced to a universal scale or fixed typology. Researchers first identify and compare confirmed episodes and pathways, then establish human-coder reliability and construct validity, and subsequently test when pathway variation explains endpoint or downstream differences such as robustness, oversight, burden, repair, agency, or governance consequences.

## 8.7. Operational calibration of the frozen identification protocol

To test the operational tractability of the proposed identification architecture, a frozen construct specification and frozen operational coding protocol were applied to three previously unseen natural human–AI conversation records. The calibration did not estimate prevalence or construct validity; it tested whether the frozen identification protocol could be applied consistently across heterogeneous natural records without adding new construct rules during execution. Table 3 summarises two separately executed model-based evaluator series across all three records.

The records deliberately placed different demands on the procedure: an extended music-creation and release workflow, a multimodal cover-image generation interaction containing refusals and technical failures, and a 97-page website-construction interaction containing uploads, code artifacts, interface pastes, external platform messages, repeated repairs, and long active task states. The same frozen identification protocol was applied unchanged across all three cases.

**Table 3. Cross-evaluator operational calibration summary**

| Case | Evaluator | Moves / candidates | Boundaries S / A / I | Episodes C / N / I | Pathways |
|---|---|---|---|---|---|
| V1 | ChatGPT | 191 / 189 | 180 / 3 / 7 | 170 / 6 / 13 | 9 |
| V1 | Claude | 192 / 188 | 171 / 0 / 19 | 152 / 0 / 36 | 14 |
| V2 | ChatGPT | 31 / 27 | 27 / 0 / 2 | 23 / 0 / 4 | 3 |
| V2 | Claude | 31 / 27 | 25 / 0 / 4 | 20 / 0 / 7 | 3 |
| V3 | ChatGPT | 286 / 284 | 277 / 2 / 6 | 268 / 4 / 12 | 8 |
| V3 | Claude | 286 / 284 | 272 / 1 / 12 | 258 / 2 / 24 | 12 |

Note. S = Supported; A = Absent; I = Insufficient; C = Confirmed; N = Non-sociodual. V1 differed by one local multimodal unitisation decision. V2 and V3 matched exactly on move and candidate counts. Remaining differences were concentrated in contingency assessment and pathway reconstruction.

Calibration showed that the core architecture was operationally tractable across heterogeneous natural records, including records containing refusals, technical failures, errors, off-target outputs, and multimodal material. Episode-status and pathway reconstruction remained governed by the frozen protocol. V2 and V3 produced exact agreement on move and candidate counts across evaluators, while V1 exposed a single local multimodal unitisation edge case rather than competing reconstructions of the interaction.

Residual disagreement was concentrated rather than diffuse. The recurrent judgment-sensitive area was return-contingency assessment at stage transitions. Additional friction was local to interleaved/multimodal source formats and to a small number of secondary orientation labels. These issues were handled within the frozen protocol without altering the construct definition or primary episode rule. Calibration therefore supports the operational coherence of the frozen architecture with limited edge-case ambiguity; across these executions, no recurring issue necessitated modification of the construct definition or primary coding rules.

## 9. Research Agenda

The research agenda should remain focused on questions that follow directly from the construct rather than treating every form of human–AI interaction as a socioduality problem. The central task is not merely to identify whether responses occurred or whether a final decision changed, but to compare how subsequent moves were formed in relation to prior exchange and how different pathways unfolded.

### 9.1. Establishing human-coder reliability and empirical validity

The frozen protocol has passed an initial model-based operational calibration, but formal human-coder reliability and empirical construct validation remain outstanding. The next priority is therefore to test whether trained independent human coders can reproduce episode identification and pathway reconstruction across domains and data channels. Reliability testing should distinguish primary evidentiary disagreements from their downstream effects on episode and pathway characterisation. A second priority is construct validity. The supplementary exploratory analysis provides initial comparative evidence that episode/pathway structure is empirically non-equivalent to blind developmental/task segmentation and is not reducible to several adjacent interaction properties. Future studies should extend this evidence through independent human coding, additional comparators, and outcome-linked validation designs. Established advice-taking measures such as WOA can complement, rather than substitute for, episode/pathway coding. Response orientation and substantive contribution re-formation should remain post-confirmation properties rather than evidence used to manufacture episode existence.

### 9.2. Comparing episode composition and pathway development

Once episodes can be identified reliably, studies can compare how different early response orientations are associated with later pathway development: continuation versus termination, verification cascades, repair, repeated disagreement, redirection, reaffirmation, or other pathway configurations. The first question is process-internal: how does early episode organisation relate to later pathway structure? Such work can test Proposition 2 without requiring a downstream performance difference. A second stage can then test whether pathway differences are associated with endpoint states or downstream outcomes such as accuracy, reproducibility, robustness, oversight, or burden.

### 9.3. Tracing information and error

Future studies should trace how correct information, errors, uncertainty, and assumptions travel across turns. Important questions include where an error first appeared, whether the other party detected it, whether a correction persisted, and whether a small distortion was amplified through reciprocal updating.

### 9.4. Examining history and context

The same immediate response may be processed differently under different histories. Research should examine which earlier elements remain active, how histories of trust or error affect later interpretation, and how newly introduced contextual inputs redirect an established interaction pathway.

### 9.5. Asymmetry, agency, and responsibility

Human and AI parties are analytically distinguishable but not ontologically equivalent. Studies should examine differences in access to history, capacity to interpret context, formal authority, and responsibility. A central governance question is whether nominal human approval reflects meaningful evaluation after the human's judgement has already been reshaped through prior AI responses.

### 9.6. Scope beyond human–AI interaction

Broader portability should remain a secondary research question. Applications beyond human–AI interaction should be considered only where a comparable process of reciprocal, response-conditioned move formation can be demonstrated and where pathway variation performs comparable analytical or explanatory work. The present article does not treat general interpersonal or institutional interaction as already established instances of the construct.

> The socioduality research agenda places the relational process first. It asks what happened between the parties, how successive moves were formed in relation to prior exchange, and how confirmed episodes and pathways differ even when endpoints do not. Only after the episode or pathway has been characterised does it ask whether those process differences also help explain outcome variation, robustness, agency, burden, repair, governance, or other consequences.

## 10. Discussion and Conclusion

### 10.1. Theoretical contribution

Socioduality establishes a bounded process-level analytical object for human–AI interaction: the conditionally connected relational sequence through which successive observable moves are formed in relation to prior exchange. Its contribution is the integration of nested units, two explicit contingency tests, evidentiary states, pathway continuity, and the separation of process characterisation from endpoint states and downstream outcomes. This architecture makes construct existence operationally inspectable before response orientation or substantive contribution re-formation is coded.

The construct is grounded in established traditions without collapsing into any one of them. Process theory, sequence analysis, grounding and repair, feedback, IRF/IRE, JAS/advice-taking, reciprocal learning, mutual adaptation, and related human–AI frameworks supply intellectual ancestry and methodological resources. Socioduality contributes the integrated episode/pathway object that these traditions illuminate from different directions: a common unit and evidence architecture for identifying reciprocal, history-carrying formation across human–AI interaction. Its distinctiveness is therefore substantive and testable rather than dependent on terminological novelty.

## 10.2. Pathway irreducibility, endpoint-state equivalence, and process difference

Pathway irreducibility is central to this contribution and is informational before it is predictive. Endpoint-state information is a many-to-one compression of the relational process record: different confirmed pathways can terminate in the same observable state, so the endpoint cannot uniquely reconstruct which pathway occurred once sequence information is discarded. This remains true whether or not those pathways later differ in performance or robustness. The WOA = 0 illustration makes the point concretely: explicit rejection after considering advice and an unchanged judgement with no observable uptake can produce the same numerical endpoint while representing different evidentiary process states. Human–AI interactions can likewise reach the same judgement through different confirmed pathways involving acceptance, rejection, verification, correction, repair, reaffirmation, error propagation, or redirection.

This distinction prevents a common compression of human–AI interaction into a start state and an end state. A favourable endpoint state can emerge from a pathway that is prolonged, conflictual, error-laden, verification-intensive, or uneven in burden; an unfavourable endpoint can follow a comparatively careful pathway. These possibilities do not create new "process quality" and "outcome quality" constructs within socioduality. They show why neither can be inferred from the other and why the pathway supports questions that disappear when only terminal state is retained.

This distinction also cautions against interpreting the final presence of a human as sufficient evidence of meaningful oversight. A human may formally approve an outcome after earlier AI responses have redirected the judgement process; alternatively, the human may explicitly consider and reject the AI while retaining the original decision. Conversely, AI influence does not automatically imply passivity. Process-sensitive governance therefore requires evidence about how the decision was formed, not only who formally signed it or whether the final answer changed.

### 10.3. Initial domain and broader portability

Human–AI interaction is the bounded domain of the present theorisation because prompts, responses, judgements, revisions, verification requests, decisions, and some resulting actions can often be observed directly or preserved chronologically. The current construct specification does not classify AI–AI dyads, non-human dyads, or genuinely triadic/network interaction. Potential portability to those settings, or to broader interpersonal and institutional contexts, requires separate theoretical and empirical examination rather than inference from the generic party definition.

### 10.4. Limitations

Three limitations define the present evidentiary boundary without weakening the construct claim. First, the calibration establishes operational tractability rather than construct validity, prevalence, predictive validity, or formal human inter-rater reliability; those require trained human coders and independent empirical tests. Second, residual coding discretion is concentrated at some return-contingency boundaries and a small number of multimodal or interleaved source-format cases, while the closed orientation vocabulary used in the calibration is occasionally coarse at the post-confirmation layer. Third, the present theorisation is bounded to observable human–AI dyads; sparse records may remain indeterminate, observational evidence cannot always establish causal dependence, and broader dyadic, triadic, or network portability must be tested separately. These limitations concern evidence, scope, and future validation. They do not alter the specified episode/pathway architecture, the two contingency requirements, or the analytical separation between process and endpoint that the frozen protocol successfully operationalised.

## 10.5. Conclusion

Socioduality identifies a process that endpoint-centred human–AI analysis systematically leaves unresolved: how one party's response becomes part of the conditions under which the other party's next observable move is formed. The construct makes that process empirically tractable through a bounded architecture of moves, candidate episodes, two relational contingencies, confirmed episodes, and linked sociodual pathways. It thereby distinguishes genuine reciprocal formation from mere temporal succession, parallel contribution, fixed alternation, and evidential uncertainty.

The resulting shift is consequential. Sociodual analysis asks not only what the human and AI produced, but which confirmed episodes occurred, how they linked into a pathway, and how successive moves were formed in relation to prior exchange. Endpoint equivalence does not imply pathway equivalence. Calibration across heterogeneous natural interactions shows that the frozen architecture can be applied without ad hoc rule creation and that residual ambiguity is localised rather than structural. The supplementary comparative analysis further shows that, in the examined natural records, Sociodual pathway topology is empirically non-equivalent to blind developmental/task segmentation, including under finer-grained segmentation and prospective unseen-case extension. Socioduality therefore provides a rigorous process-level basis for studying human–AI interaction as interaction, preserving relational information that endpoint-centred approaches cannot recover.

## Generative AI Disclosure

Generative AI tools were used under the author's direction for language editing, stylistic refinement, formal formatting, the model-based operational calibration reported in Section 8.7, and the model-based blind analyses reported in the Supplementary Exploratory Analysis. The conceptual framework, protocol design, study decisions, analytic comparison rules, interpretation of findings, and final content were developed and determined by the author.

## Supplementary Material

A Supplementary Exploratory Analysis follows the References. It compares frozen Sociodual pathway reconstructions with blind developmental/task-process segmentations, reports granularity and record-format robustness checks, and extends the comparison to three prospectively selected unseen natural human–AI interactions. The supplementary analysis does not alter the construct specification.

## References

Bailey, P. E., Leon, T., Ebner, N. C., Moustafa, A. A., & Weidemann, G. (2023). A meta-analysis of the weight of advice in decision-making. Current Psychology, 42(28), 24516–24541. doi:10.1007/s12144-022-03573-2

Bonaccio, S., & Dalal, R. S. (2006). Advice taking and decision-making: An integrative literature review, and implications for the organizational sciences. Organizational Behavior and Human Decision Processes, 101(2), 127–151. doi:10.1016/j.obhdp.2006.07.001

Clark, H. H., & Brennan, S. E. (1991). Grounding in communication. In L. B. Resnick, J. M. Levine, & S. D. Teasley (Eds.), Perspectives on socially shared cognition (pp. 127–149). American Psychological Association. doi:10.1037/10096-006

Çögenli, M. Z. (2026). Visible AI assistant outputs in psychosocial risk management: An OHP/OHS-grounded benchmark for governance and responsible use. Frontiers in Public Health, 14, 1857113. doi:10.3389/fpubh.2026.1857113

Davis, N. (2026). Interaction-Centered Intelligence: Toward an interaction-based theory of human–AI co-creation [Preprint]. arXiv. doi:10.48550/arXiv.2606.00807

Ding, S., & Tan, S. (2026). Evaluating multi-turn human–AI interaction. In Proceedings of the Workshop on Evaluating Evaluations (EvalEval) (pp. 12–18). Association for Computational Linguistics. doi:10.18653/v1/2026.evaleval-1.2

Giddens, A. (1984). The Constitution of Society: Outline of the Theory of Structuration. University of California Press.

Glickman, M., & Sharot, T. (2025). How human–AI feedback loops alter human perceptual, emotional and social judgements. Nature Human Behaviour, 9, 345–359. doi:10.1038/s41562-024-02077-2

Johnson, M., Bradshaw, J. M., Feltovich, P. J., Jonker, C. M., van Riemsdijk, M. B., & Sierhuis, M. (2014). Coactive design: Designing support for interdependence in joint activity. Journal of Human-Robot Interaction, 3(1), 43–69. doi:10.5898/JHRI.3.1.Johnson

Langley, A. (1999). Strategies for theorizing from process data. Academy of Management Review, 24(4), 691–710. doi:10.5465/amr.1999.2553248

Mehan, H. (1979). Learning Lessons: Social Organization in the Classroom. Harvard University Press.

Nikolaidis, S., Kuznetsov, A., Hsu, D., & Srinivasa, S. (2016). Formalizing human–robot mutual adaptation: A bounded memory model. In Proceedings of the 11th ACM/IEEE International Conference on Human-Robot Interaction (pp. 75–82). doi:10.1109/HRI.2016.7451736

Pedreschi, D., Pappalardo, L., Ferragina, E., Baeza-Yates, R., Barabási, A.-L., Dignum, F., Dignum, V., Eliassi-Rad, T., Giannotti, F., Kertész, J., Knott, A., Ioannidis, Y., Lukowicz, P., Passarella, A., Pentland, A. S., Shawe-Taylor, J., & Vespignani, A. (2025). Human–AI coevolution. Artificial Intelligence, 339, Article 104244. doi:10.1016/j.artint.2024.104244

Sacks, H., Schegloff, E. A., & Jefferson, G. (1974). A simplest systematics for the organization of turn-taking for conversation. Language, 50(4), 696–735. doi:10.2307/412243

Schegloff, E. A. (2007). Sequence Organization in Interaction: A Primer in Conversation Analysis, Volume 1. Cambridge University Press. doi:10.1017/CBO9780511791208

Schegloff, E. A., Jefferson, G., & Sacks, H. (1977). The preference for self-correction in the organization of repair in conversation. Language, 53(2), 361–382. doi:10.2307/413107

Sinclair, J. M., & Coulthard, R. M. (1975). Towards an Analysis of Discourse: The English Used by Teachers and Pupils. Oxford University Press.

Sniezek, J. A., & Buckley, T. (1995). Cueing and cognitive conflict in judge–advisor decision making. Organizational Behavior and Human Decision Processes, 62(2), 159–174. doi:10.1006/obhd.1995.1040

Suchman, L. A. (1987). Plans and Situated Actions: The Problem of Human-Machine Communication. Cambridge University Press.

Suddaby, R. (2010). Editor's comments: Construct clarity in theories of management and organization. Academy of Management Review, 35(3), 346–357. doi:10.5465/amr.35.3.zok346

Sydow, J., Schreyögg, G., & Koch, J. (2009). Organizational path dependence: Opening the black box. Academy of Management Review, 34(4), 689–709. doi:10.5465/amr.34.4.zok689

Te'eni, D., Yahav, I., Zagalsky, A., Schwartz, D. G., Silverman, G., Cohen, D., Mann, Y., & Lewinsky, D. (2023). Reciprocal human–machine learning: A theory and an instantiation for the case of message classification. Management Science, 72(1), 167–192. doi:10.1287/mnsc.2022.03518

Vaccaro, M., Almaatouq, A., & Malone, T. W. (2024). When combinations of humans and AI are useful: A systematic review and meta-analysis. Nature Human Behaviour, 8, 2293–2303. doi:10.1038/s41562-024-02024-1

Van de Ven, A. H., & Poole, M. S. (1995). Explaining development and change in organizations. Academy of Management Review, 20(3), 510–540. doi:10.5465/amr.1995.9508080329

Wang, X., Wang, Z., Liu, J., Chen, Y., Yuan, L., Peng, H., & Ji, H. (2024). MINT: Evaluating LLMs in multi-turn interaction with tools and language feedback. In Proceedings of the 12th International Conference on Learning Representations (ICLR 2024).

Wiener, N. (2019). Cybernetics or Control and Communication in the Animal and the Machine (reissue of the 1961 second edition; original work published 1948). MIT Press. doi:10.7551/mitpress/11810.001.0001

# Supplementary Exploratory Analysis (*Socioduality*): Blind Comparison of Sociodual Pathway Individuation and Developmental Process Segmentation

*Supplement to: "Socioduality: A Relational Process Framework for Human-AI Interaction"*

Mehmed Zahid Çögenli, PhD
Department of Occupational Health and Safety, Faculty of Health Sciences
Uşak University, Uşak, Türkiye
mzahid.cogenli@usak.edu.tr | ORCID: 0000-0003-3018-4157

**Case-protection note.** The six natural interaction records are reported only as D1-D3 and U1-U3. Case labels are deliberately non-descriptive. Content examples are functionally abstracted where literal wording is not required to audit the analysis. Anonymisation does not alter move order, boundary placement, episode status, pathway ranges, or quantitative results.

## 1. Purpose and Analytic Logic

This supplementary exploratory analysis asks whether Sociodual pathway reconstruction merely reproduces ordinary developmental/task-process segmentation or preserves a different relational structure. Ordinary process analysis identifies changes in what the interaction is doing. Socioduality instead identifies whether successive contributions remain connected through observable response and return contingency and therefore belong to one maximal uninterrupted relational pathway.

The comparison is discriminant rather than competitive. If the two representations repeatedly place the same boundaries, the pathway architecture may add little beyond relabelled task segmentation. If they retain and terminate continuity at different locations under frozen rules, they preserve different information about the same interaction record. The package therefore combines a discovery comparison, a record-format negative control, a fine-grained granularity stress test, and a prospective unseen extension.

## 2. Materials and Analytic Procedure

### 2.1. Analytic chronology

The Sociodual identification architecture and Coding Manual v0.2 were frozen before the supplementary comparison, and the D1-D3 Sociodual pathway maps already existed before blind ordinary developmental analysis. Later robustness analyses were allowed to challenge or qualify the comparison but not to merge, split, or repair frozen pathways.

| Stage | Analytic step | Function |
|---|---|---|
| 1 | Operational freeze | Move, boundary, candidate, episode, and maximal-pathway rules fixed. |
| 2 | Pre-existing D1-D3 maps | Independent frozen Sociodual maps existed before the comparison. |
| 3 | Blind developmental reading | Ordinary task/process analyses produced without Sociodual materials. |
| 4 | Boundary-rule freeze | Move coordinate, exact matching, ±1 sensitivity, and one-to-one matching fixed. |
| 5 | Robustness tests | Format control and fine-grained ordinary re-segmentation. |
| 6 | Prospective unseen extension | U1-U3 selected before the present outcomes were inspected; analyses kept separate. |

*Table S1. Analytic chronology. "Prospective" denotes temporal ordering and unseen case selection relative to the discovery finding; it is not presented as formal preregistration.*

### 2.2. Case sets

The discovery set comprised three heterogeneous natural human-AI records, D1-D3. These same records had previously been used for operational calibration of the frozen coding procedure. The discovery comparison is therefore not an out-of-sample test of the pathway architecture. The later U1-U3 set consisted of three complete natural conversations selected prospectively from the available archive rather than chosen because they appeared especially reciprocal or likely to support Socioduality.

### 2.3. Sociodual pathway reconstruction

Sociodual coding followed the current construct specification together with Coding Manual v0.2. Source fidelity and move unitisation preceded candidate generation; each adjacent cross-party boundary received one Supported, Absent, or Insufficient judgment; candidate status followed mechanically; secondary coding was restricted to Confirmed episodes; and pathways were reconstructed as maximal uninterrupted chains of at least two overlapping Confirmed episodes. Indeterminate or Non-sociodual candidates formally interrupted a pathway and were not bridged by topic, artifact, project, or actor continuity.

D1-D3 retain the two pre-existing model-based Sociodual evaluator maps separately. U1-U3 have one complete fresh Sociodual line through GPT, executed independently of the ordinary segmentation outputs. The unseen extension is therefore reported as a prospective GPT-line Sociodual replication, not as a two-model unseen Sociodual replication.

### 2.4. Blind ordinary process analysis

Ordinary-process readers received the natural conversation record and a theory-neutral instruction to identify substantive transitions in task, subtask, goal, problem, functional purpose, or object of work. They were not given the Sociodual construct, Coding Manual, pathway maps, comparison results, or another reader's output. A later fine-grained instruction required the finest substantively defensible segmentation without creating boundaries merely for speaker changes, wording changes, errors, corrections, failures, uploads, or modality changes when the local task continued.

### 2.5. Common boundary coordinate and comparison rules

All positional comparisons use the frozen move coordinate M_k | M_k+1. Exact positional overlap (±0) is primary; ±1 move is reported separately as a sensitivity analysis. Matching is one-to-one, with exact matches assigned before distance-one matches. A stage-only boundary is an ordinary task/developmental transition inside a continuing pathway. A pathway-only boundary is a formal pathway break not matched by an ordinary boundary. Overlap is a matched boundary. Gap-stage is an ordinary boundary located between pathways without matching a formal break anchor. The earlier 'bridged breaks' display has been removed because, in the broad discovery table, it was numerically redundant with pathway-only breaks rather than supplying an independent measure.

The primary discovery comparison in Table S2 uses the broad ChatGPT ordinary developmental reading against each of the two frozen Sociodual maps. The positional granularity and unseen audits in Tables S4 and S6 use one fixed pairing throughout: the Claude blind ordinary boundary line against the GPT Sociodual map. This keeps the move-level audit pairing constant across those two tests and avoids multiplying dependent cross-pairings. The independent ChatGPT ordinary reader remains a descriptive cross-reader check on segmentation resolution.

## 3. Discovery Results: Structural Non-Equivalence

| Record | Frozen map | Dev. boundaries | Formal breaks | Stage-only | Pathway-only | Overlap | Gap-stage |
|---|---|---|---|---|---|---|---|
| D1 | GPT | 6 | 9 | 4 | 7 | 2 | 0 |
| D1 | Claude | 6 | 14 | 2 | 11 | 3 | 1 |
| D2 | GPT | 2 | 2 | 1 | 1 | 1 | 0 |
| D2 | Claude | 2 | 2 | 1 | 1 | 1 | 0 |
| D3 | GPT | 7 | 7 | 7 | 7 | 0 | 0 |
| D3 | Claude | 7 | 12 | 7 | 12 | 0 | 0 |
| Pooled descriptive | — | 30 | 46 | 22 | 39 | 7 | 1 |

*Table S2. Primary discovery boundary comparison. The pooled row is descriptive only; the six record-by-map comparisons are not independent observations. Exact and ±1 matching produced the same classifications in all six rows.*

All six record-by-map comparisons contained both stage-only and pathway-only boundaries. Of 30 mapped developmental boundaries, 22 were stage-only, seven overlapped a formal break, and one fell in an inter-pathway gap. Of 46 formal breaks, 39 were pathway-only and seven overlapped a developmental boundary. D3 is particularly diagnostic: both frozen maps produced zero overlap (GPT: 7 stage-only, 7 pathway-only, 0 overlap; Claude: 7, 12, 0), and the zero remained under ±1 sensitivity. It is therefore not an artifact of requiring exact same-index matches.

Because D1-D3 are also the calibration records, these discovery results are evidence of structural non-equivalence within the records on which the operational procedure had already been exercised, not an independent generalisation sample. The prospective U1-U3 extension in Section 5 was included specifically to reduce that dependence.

A Sociodual pathway can span multiple developmental/task stages, and a single developmental/task context can contain multiple formal Sociodual pathways.

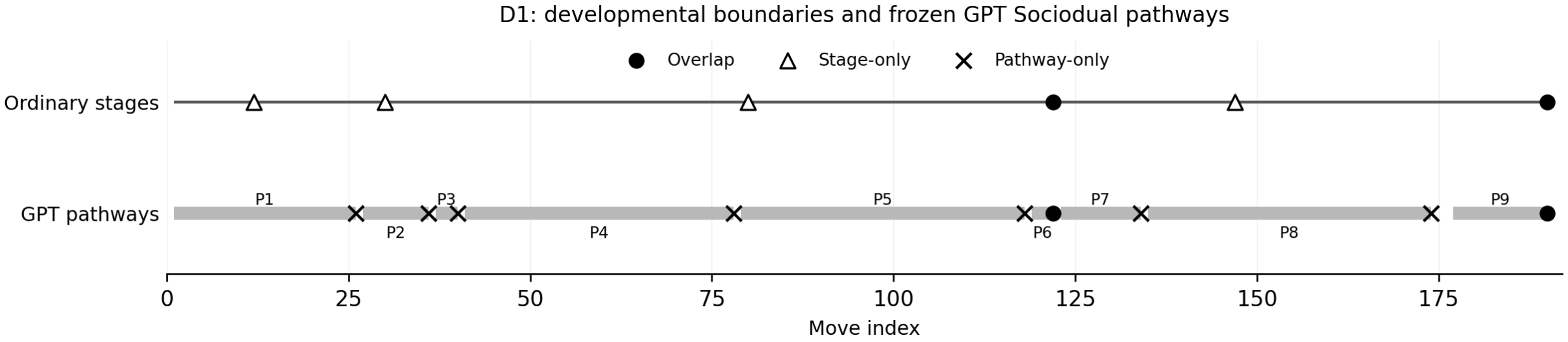

*Figure S1. D1 move-level display using the frozen GPT map for legibility. Filled circles indicate overlap, triangles stage-only boundaries, and x marks pathway-only breaks. All nine GPT pathways are labelled. The rightmost overlap is M190 | M191; M191 is an observed subsequent move, so this marker is not a record-end anchor. Table S2 reports both frozen evaluator maps.*

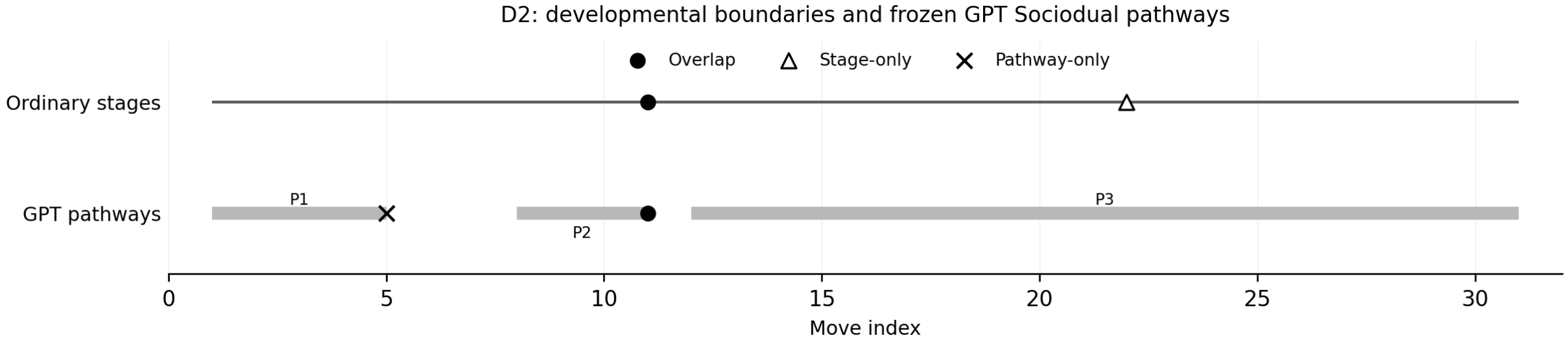

*Figure S2. D2 move-level display using the frozen GPT map. The same three-marker scheme distinguishes overlap from non-overlap directly.*

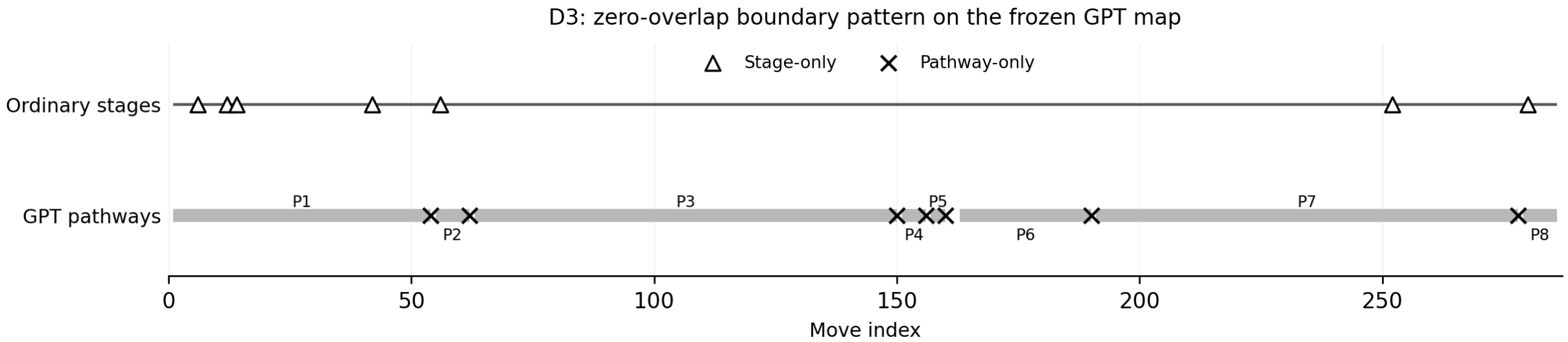

*Figure S3. D3 move-level display using the frozen GPT map. No filled-circle overlap occurs: all seven developmental boundaries are stage-only and all seven formal breaks are pathway-only. The zero-overlap result is unchanged under ±1 sensitivity; the separate Claude map likewise has zero overlap (7 stage-only, 12 pathway-only).*

# 4. Robustness Analyses

## 4.1. Record-format negative control

The negative control used the GPT discovery move ledgers and the corresponding frozen GPT pathways as one fixed reference. A format transition was counted conservatively when adjacent ledger entries switched between plain text-only and an explicitly marked enriched or structured modality (for example, attachment/image, generated artifact, code, structured selection, interface text, quoted external material, or URL-bearing contribution). Material that an evaluator had left coded simply as Text was not reclassified from prose content alone. The question was whether formal breaks clustered near these format changes and whether format changes themselves forced breaks.

| Record | GPT formal breaks | Format transitions | Format-transition density | Breaks within ±1 of format transition | Format transitions inside a pathway |
|---|---|---|---|---|---|
| D1 | 9 | 4 | 4 / 190 (2.1%) | 1 / 9 (11.1%) | 4 / 4 |
| D2 | 2 | 10 | 10 / 30 (33.3%) | 1 / 2 | 9 / 10 |
| D3 | 7 | 82 | 82 / 285 (28.8%) | 3 / 7 (42.9%) | 79 / 82 |
| Pooled descriptive | 18 | 96 | 96 / 505 (19.0%) | 5 / 18 (27.8%) | 92 / 96 |

*Table S3. Quantitative record-format negative control. Format-transition density is the number of ledger-explicit format-transition boundaries divided by all adjacent move boundaries in the record. Counts are descriptive and use ledger-explicit modality changes rather than inferred content-format categories.*

Only 5 of 18 formal breaks were within ±1 move of a ledger-explicit format transition, while 92 of 96 format-transition boundaries occurred inside an uninterrupted pathway. The raw format-transition density was 2.1% in D1, 33.3% in D2, 28.8% in D3, and 19.0% pooled. Because the break criterion allows a ±1 window, raw density is a conservative exposure baseline rather than the direct expected match rate. Under a simple independence approximation, $q = 1 - (1 - p)^3$, weighted by each record's number of formal breaks, approximately 35.7% of breaks would be expected to fall within ±1 of a format transition by chance, compared with 27.8% observed. This benchmark is descriptive rather than an inferential significance test. Format change was therefore neither necessary for most breaks nor sufficient to generate a break in the discovery records.

## 4.2. Fine-grained granularity stress test

Fresh blind fine-grained readers increased ordinary process resolution: ChatGPT produced 49, 14, and 38 segments for D1-D3; Claude produced 64, 13, and 76. To test convergence with the discovery metric, move-mappable Claude boundaries were compared with the frozen GPT formal-break anchors under the same exact and ±1 rules.

| Record | Ordinary boundaries | Move-mappable | Within-turn | Formal breaks | Stage-only | Pathway-only | Overlap | Gap-stage | ±1: S / P / O / G |
|---|---|---|---|---|---|---|---|---|---|
| D1 | 63 | 60 | 3 | 9 | 50 | 0 | 9 | 1 | 50 / 0 / 9 / 1 |
| D2 | 12 | 12 | 0 | 2 | 9 | 0 | 2 | 1 | 9 / 0 / 2 / 1 |
| D3 | 75 | 73 | 2 | 7 | 66 | 1 | 6 | 1 | 66 / 1 / 6 / 1 |
| Pooled descriptive | 150 | 145 | 5 | 18 | 125 | 1 | 17 | 3 | 125 / 1 / 17 / 3 |

*Table S4. Fine-grained granularity stress test using the same frozen boundary metric. D1-D3 exact classifications were unchanged under ±1 sensitivity. "Within-turn" boundaries were retained as ordinary segmentation results but are not forced onto an inter-move coordinate.*

Fine-graining recovered 17 of 18 GPT formal breaks exactly while leaving 125 move-mappable ordinary boundaries inside uninterrupted pathways; one break remained unmatched and three ordinary boundaries fell in inter-pathway gaps. Higher resolution therefore increased local overlap without making the two topologies identical.

Fine-grained ordinary process segmentation does not collapse into the frozen Sociodual pathway topology; greater task resolution changes overlap density without making the boundary systems identical.

## 5. Prospective Unseen Replication

Three prospectively selected unseen natural conversations were analysed without altering the construct or coding rules. Fresh GPT Sociodual executions produced valid move ledgers and pathways in all three; independent blind fine-grained ordinary analyses were produced by ChatGPT and Claude.

| Record | Moves | Confirmed | Non-sociodual | Indeterminate | Max. pathways | Confirmed singletons | Re-formation: Present / Absent / Insufficient |
|---|---|---|---|---|---|---|---|
| U1 | 76 | 66 | 0 | 8 | 5 | 0 | 64 / 2 / 0 |
| U2 | 104 | 80 | 4 | 0 | 7 | 4 | 72 / 8 / 0 |
| U3 | 58 | 52 | 0 | 2 | 2 | 0 | 48 / 0 / 4 |

*Table S5. Fresh GPT Sociodual results for unseen records. Confirmed singletons are shown separately because one Confirmed episode is not a pathway.*

| Record | Ord. boundaries | Compared | Within-turn | Formal breaks | Stage-only | Pathway-only | Overlap | Gap-stage | ±1: S / P / O / G | Sensitivity |
|---|---|---|---|---|---|---|---|---|---|---|
| U1 | 37 | 37 | 0 | 4 | 33 | 0 | 4 | 0 | 33 / 0 / 4 / 0 | No change |
| U2 | 43 | 43 | 0 | 7 | 30 | 2 | 5 | 8 | 29 / 1 / 6 / 8 | P-only → overlap |
| U3 | 30 | 27 | 2 | 1 | 27 | 1 | 0 | 0 | 26 / 0 / 1 / 0 | P-only → overlap |

*Table S6. Prospective unseen boundary comparison under the same frozen rules as Table S4. For U3, one additional inter-move ordinary boundary precedes the first eligible Sociodual pathway interval and is therefore outside the four-way comparison; two further boundaries are within-turn.*

The central non-equivalence finding replicated in all three unseen records on the fresh GPT Sociodual line. U1 shows extensive stage-only structure with all four formal breaks recovered by the blind ordinary line. U2 is bidirectionally non-equivalent even after ±1 sensitivity: 29 stage-only boundaries and one pathway-only break remain, with six overlaps. U3 changes locally under ±1 because its single formal break lies one move from an ordinary boundary; even then, 26 ordinary transitions remain inside the long continuing pathway. The mismatch form therefore varies by interaction, but none of the three unseen records collapses to identical ordinary and Sociodual topology.

## 6. Cross-Case Analytical Distinctiveness

| Comparator | Observed distinction | Implication |
| --- | --- | --- |
| Developmental/task segmentation | Stage transitions occur inside pathways; formal breaks also occur inside continuing task contexts. | Not a relabeling of task progression. |
| Analytic granularity | Fine-graining recovered many formal breaks but left 125 move-mappable task transitions inside pathways in D1-D3. | Not a simple zoom-level effect. |
| Substantive re-formation | U1: 2 Confirmed RF-Absent; U2: 8 Confirmed RF-Absent. | Relational conditioning can exist without substantive adaptation. |
| Agreement / success | Confirmed pathways contain rejection, redirection, verification, implementation failure, error, and correction. | Not a success or agreement label. |
| Topic continuity | Task/topic change can occur within a pathway; continued task/topic does not bridge a formal break. | Topic identity does not determine pathway continuity. |
| Record format / modality | 5/18 breaks near format transition; 92/96 format transitions inside pathways. | Topology is not reducible to record format. |
| Whole-session membership | A session can contain multiple pathways, breaks, Non-sociodual or Indeterminate candidates, and Confirmed singletons. | The interaction container is not itself the pathway unit. |

*Table S7. Multi-axis analytical distinctiveness using the existing coded evidence package.*

Across these axes, the pathway construct is not reducible to a second coding of task progression, adaptation, agreement, success, topic continuity, modality, or session membership. The re-formation counts are especially useful because they separate construct existence from substantive change: U1 contains two and U2 eight Confirmed episodes in which substantive re-formation was Absent, while reciprocal contingency was still supported. The remaining rows show the complementary point that rejection, failure, topic shifts, and format shifts can all occur without determining pathway existence or termination.

## 7. Interaction History: Continuous Carryover and Post-Break Reuse

The history-carrying premise does not imply that every later use of earlier material belongs to one uninterrupted pathway. Existing records show two observable situations without requiring a new formal subtype: earlier states may remain consequential within a continuing Sociodual pathway, or an earlier decision, artifact, name, or constraint may become consequential again after a formal break. In the latter case, later relevance does not retroactively restore the interrupted contingency chain.

| Record | Earlier product/state | Later consequence | Interpretation |
| --- | --- | --- | --- |
| D1 | An earlier relationally formed identity/label | Reappears later after a formal interruption | Post-break history reuse; the break remains formal. |
| D2 | Shared design logic and specification | Remains consequential through planning, production failure, negotiation, and workaround | Continuous carryover inside one pathway. |

| Record | Earlier product/state | Later consequence | Interpretation |
| --- | --- | --- | --- |
| D3 | Earlier infrastructure constraint | Becomes consequential when a later alternative infrastructure solution is considered | Later history reuse does not erase the intervening break. |

*Table S8. Existing-data illustrations of history within and across formal pathway boundaries.*

Interaction history can remain consequential without implying uninterrupted relational continuity.

## 8. Worked Example: D2 at Move Level

| Representation | Boundary | Functional location | Status |
| --- | --- | --- | --- |
| Developmental | M011 \| M012 | Broad stage transition | Overlap with a formal break. |
| Developmental | M022 \| M023 | Planning/specification to execution/failure transition | Stage-only; later pathway continues. |
| GPT frozen map | M005 \| M006 | Formal interruption within the first broad developmental stage | Pathway-only. |
| GPT frozen map | M011 \| M012 | End of P02 | Overlap. |
| Claude frozen map | M004 \| M005 | Formal interruption within the first broad developmental stage | Pathway-only. |
| Claude frozen map | M011 \| M012 | End of P02 | Overlap. |

*Table S9. D2 primary boundary audit. GPT pathways are M001-M005, M008-M011, and M012-M031; Claude pathways are M001-M004, M008-M011, and M014-M031.*

D2 contains both directions in compact form. The early formal interruption occurs while the blind developmental account remains in the same broad work phase, so task continuity does not license relational continuity. Later, M014-M031 remains one uninterrupted pathway while the work moves through source input, concept selection, specification, production attempt, failure, responsibility negotiation, and workaround. Production failure itself does not sever the pathway when successive contributions remain observably contingent on one another.

## 9. Interpretation, Limitations, and Empirical Contribution

### 9.1. Interpretation

The supplementary evidence supports treating reciprocal evidentiary continuity as a process dimension distinct from ordinary task progression in the examined records. An interaction can change task while remaining one relational formation, or continue the same task after that formation has formally broken. Pathway reconstruction therefore preserves relational continuity information that task stage, output, or endpoint descriptions do not uniquely recover.

### 9.2. Limitations

The evidence is deliberately bounded. First, the package contains six natural interaction records; it does not estimate population prevalence. Second, D1-D3 are also the records used in operational calibration, so their discovery result is not out-of-sample; U1-U3 reduce but do not erase that calibration dependence. Third, all evaluators in the present package are model-based, and no formal human-coder reliability estimate is available. Fourth, the integrating analyst performed the move-coordinate mapping and descriptive count calculations rather than using an independent human adjudication panel. Fifth, the unseen Sociodual side contains one complete fresh GPT line, although the ordinary unseen side has both ChatGPT and Claude blind readings. Sixth, some records contain local source-fidelity or unitisation limitations, which were retained rather than reconstructed. Finally, the analyses do not test predictive validity, causal effects, downstream performance, trust, agency, robustness, or other outcomes.

### 9.3. Overall empirical contribution

Across six heterogeneous natural human-AI interactions, Sociodual pathway reconstruction produced an interaction topology empirically non-equivalent to blind developmental/task process segmentation. In the discovery set, non-equivalence occurred in both directions and remained unchanged under ±1 boundary tolerance. Fine-grained re-segmentation increased local overlap but left extensive stage-only structure and did not collapse the topologies. Record-format variation did not account for the pathway breaks. The central non-equivalence finding then replicated in three prospectively selected unseen cases on the fresh GPT Sociodual coding line, although the exact form of mismatch varied across records. For HCI research, this supplies a process-level unit for examining how human and AI contributions remain relationally linked across time rather than inferring continuity from task stage or endpoint alone.

Socioduality performs distinct analytical work in the examined records: it individuates evidence-bounded relational formations according to reciprocal evidentiary continuity, preserving process structure that ordinary task progression does not uniquely recover.

## 10. Documentation and Reproducibility

### 10.1. Audit identifiers and provenance

Auditability is preserved through move indices, developmental/task boundary indices, pathway ranges, formal-break anchors, candidate/pathway summary statistics, exact blind instructions, analytic chronology, case-selection provenance, worked examples, and comparison figures. Literal content is paraphrased or functionally abstracted when exact wording is not required to verify a structural judgment.

### 10.2. Separation of analytic inputs

Ordinary-process readers received only the assigned conversation record and the blind process-analysis instruction; they did not receive the Sociodual construct specification, Coding Manual, pathway maps, comparison rules, prior findings, or another reader's output. The D1-D3 Sociodual maps predated the comparative blind readings. Fresh U1-U3 Sociodual executions were conducted separately from the ordinary segmentations.

### 10.3. Evaluator lines

The discovery package retains separate frozen GPT and Claude Sociodual maps rather than a consensus map. The fine-grained and unseen ordinary-process conditions include independent ChatGPT and Claude readers. The unseen Sociodual extension contains one complete fresh GPT line. Exact backend build identifiers are not inferred where the archived execution record preserves only product identity.

### 10.4. Frozen analytic units

The quantitative package keeps the following units distinct:

- observable moves
- adjacent cross-party boundaries
- eligible alternating three-move candidate episodes
- Confirmed / Non-sociodual / Indeterminate episode statuses
- Confirmed singletons
- maximal uninterrupted Sociodual pathways
- ordinary process segments
- ordinary process boundaries

- formal pathway-break anchors

Primary discovery boundary classifications were frozen after the comparison rules were applied. The same exact and ±1 matching logic was then used in the granularity and unseen boundary audits; no pathway was recoded in response to those later results.

## Appendix A. Blind Analysis Instructions

### A1. Fine-grained developmental/task segmentation prompt

Independent Fine-Grained Process Segmentation — Blind Reading

Read the attached complete human–AI conversation record carefully from beginning to end.

Treat it as a natural interaction record. You have not been given, and should not assume, any particular theoretical framework, coding scheme, expected finding, or prior analysis.

Your task is not to summarise the conversation broadly. Analyse its development at a fine-grained task/process level.

Segment the record into the finest substantively defensible sequence of task, subtask, goal, problem, or work-phase units.

Identify every meaningful local transition at which the interaction begins doing something observably different from what it was doing immediately before.

Do not create a new segment merely because the speaker changes; a new message begins; wording or tone changes; an error, correction, disagreement, success, or failure occurs while the same local task continues; or a file, image, code block, interface element, or other modality appears without a corresponding substantive task/process transition.

For each segment, report its first observable contribution, last observable contribution, and the local task/subtask/goal/problem defining the segment. Then provide a chronological boundary list identifying the transition between every pair of adjacent segments and briefly state the observable basis for placing that boundary. Where the exact location is genuinely ambiguous, give the narrowest defensible range rather than forcing false precision. Do not force a predetermined number of segments. Base every judgment only on the supplied conversation record.

### A2. Independent Socioduality single-case prompt for unseen executions

Independent Socioduality Single-Case Analysis — Blind Execution

Read the supplied current Socioduality construct specification and Coding Manual v0.2 completely before analysing the conversation.

Then analyse the entire supplied natural human–AI conversation from its actual beginning to its actual end using the frozen Socioduality identification architecture and the operational rules in Coding Manual v0.2.

You have not been given, and must not infer or assume, any prior analysis of this conversation, developmental/task segmentation, previous evaluator output, comparison result, or expected finding.

Apply the supplied rules as written. Do not revise, extend, simplify, or create new coding rules or categories.

Produce a complete auditable single-case analysis including the move ledger; eligible candidate episodes; response-contingency and return-contingency judgments; Confirmed / Non-sociodual / Indeterminate classifications; secondary coding where applicable; maximal uninterrupted Sociodual pathways with exact move ranges; and a concise case-level summary.

Do not compare the resulting Sociodual pathways with task stages, developmental phases, topics, or any other segmentation system. Base every judgment only on the supplied materials and the observable conversation record.

## Appendix B. Source and Version Note

This Supplement supports the accompanying manuscript, Socioduality: A Relational Process Framework for Human-AI Interaction. Coding Manual v0.2 is used only as the frozen operational companion that generated the underlying move, boundary, candidate, episode, and pathway decisions.

Its historical reference to an earlier construct version documents the chronology of the freeze; no new coding rule, formal category, or pathway subtype was introduced for this supplementary comparison.